\font\ninerm=cmr9  \font\sixrm=cmr6
\font\ninei=cmmi9  \font\sixi=cmmi6
\font\ninesy=cmsy9  \font\sixsy=cmsy6

\font\ninebf=cmbx9  \font\sixbf=cmbx6

\font\twelverm=cmr12 \font\twelvei=cmmi12 \font\twelvesy=cmsy12
\font\twelveit=cmti12 \font\twelvesl=cmsl12 \font\twelvebf=cmbx12
\font\twelvett=cmtt12 

\font\bgp=cmbx12 scaled\magstep1

\def\twelvepoint{\def\rm{\fam0\twelverm}%
  \textfont0=\twelverm \scriptfont0=\ninerm \scriptscriptfont0=\sixrm
  \textfont1=\twelvei \scriptfont1=\ninei \scriptscriptfont1=\sixi
  \textfont2=\twelvesy \scriptfont2=\ninesy \scriptscriptfont2=\sixsy
  \textfont3=\tenex \scriptfont3=\tenex \scriptscriptfont3=\tenex
  \def\it{\fam\itfam\twelveit}%
  \textfont\itfam=\twelveit
  \def\sl{\fam\slfam\twelvesl}%
  \textfont\slfam=\twelvesl
  \def\bf{\fam\bffam\twelvebf}%
  \textfont\bffam=\twelvebf \scriptfont\bffam=\ninebf
   \scriptscriptfont\bffam=\sixbf
  \def\tt{\fam\ttfam\twelvett}%
  \textfont\ttfam=\twelvett
}


%
\twelvepoint\rm
\hsize= 6.5truein
\vsize= 8.50truein
\hoffset= 0.0truein
\voffset= 0.0truein
\lineskip= 2pt
\lineskiplimit= 2pt
\overfullrule=0pt
\tolerance=2000
\topskip= 0pt
\baselineskip=15pt
\parindent=0.4truein
\parskip=0pt plus1pt
\footline={\hss\twelverm\folio\hss}
\def\medskip{\vskip6pt plus2pt minus2pt}
\def\bigskip{\vskip12pt plus4pt minus4pt}
\def\smallskip{\vskip3pt plus1pt minus1pt}
\centerline{\bgp Electronic Structure of 
La$_{\bf 1.85}$Sr$_{\bf 0.15}$CuO$_{\bf 4}$:}
\centerline{\bgp Characterization of a Fermi Level Band Crossing}
\bigskip
\centerline{Jason K. Perry and Jamil Tahir-Kheli}
\centerline{\it First Principles Research, Inc.}
\centerline{\it 8391 Beverly Blvd., Suite \#171, Los Angeles, CA 90048}
\bigskip
\bigskip
\bigskip
\noindent{\bf Abstract:} We present the results of a new Hubbard model for 
optimally doped $La_{2-x}Sr_xCuO_4$.  This model uses parameters derived
from BLYP calculations on the cluster $CuO_6$.  It explicitly includes
the $Cu\ d_{x^2-y^2}$ and $d_{z^2}$ orbitals, the $O\ p_{\sigma}$ orbitals,
and the apical $O\ p_z$ orbitals.  When correlation is properly included
in the Hubbard model, we find that there is a crossing of two bands in
the vicinity of the Fermi level for the optimally doped superconductor.
This crossing rigorously occurs along the $(0,0)-(\pi/a,\pi/a)$ direction
of the 2-D Brillouin zone.  The crossing arises due to the overlap of a
broad ``$B_{1g}$'' band dominated by $Cu\ d_{x^2-y^2}$ character and a
narrower ``$A_{1g}$'' band dominated by $Cu\ d_{z^2}$ character.
We conclude that optimal doping of $La_{2-x}Sr_xCuO_4$ and related
materials is achieved when the Fermi level coincides with this crossing.
At this point, formation of Cooper pairs between the two bands
(i.e. inter-band pairing or IBP) leads to superconductivity.  Furthermore,
using geometric considerations, we extend our conclusions to
$YBa_2Cu_3O_{6+\delta}$ and offer a simple explanation for the seemingly
complex behavior of $T_c$ as a function of doping in this material.
This behavior can be understood on the basis of multiple band crossings.
\bigskip
\noindent \hbox to \hsize{
\hfil {\tt cond-mat/9711184}}
\bigskip
\noindent {\it Submitted to Phys. Rev. B}. 
\footnote{}{\ninerm A PostScript version of this paper is available for download
at http://www.firstprinciples.com.}
\vfill
\eject
\noindent{\bf I. Introduction.}

Eleven years have elapsed since the discovery of high-temperature copper
oxide superconductors.$^1$  In this work, we focus on the recent
suggestion by Tahir-Kheli$^2$ that superconductivity arises in
these materials due to an inter-band pairing mechanism (IBP).  That is,
Cooper pairing of electrons belonging to two distinct bands may be the
cause of superconductivity.  Such a proposal is appealing as it readily
explains why superconductivity is only observed at very specific doping
levels.  $La_{2-x}Sr_xCuO_4$, for instance, shows an optimal $T_c$
of 39 K when $x=0.15$, but superconductivity quickly vanishes as the
doping level is changed.$^3$  The IBP theory only requires that two
bands cross.  Should such a crossing exist, then optimal doping is
achieved when the Fermi level coincides with this crossing.  Precise
doping levels are needed to achieve this.

While such band crossings have not been previously noted by others,
we find a crossing to occur in optimally doped $La_{1.85}Sr_{0.15}CuO_4$
within 0.15 eV of the Fermi level using a simple Hubbard model
that includes the effect of electron correlation.  The crossing occurs
between one band which is dominated by $Cu\ d_{x^2-y^2}$ character and
a second which is dominated by $Cu\ d_{z^2}$ character.  Moreover,
we find that when the model is empirically adjusted to include missing
electronic effects, the crossing can be seen to occur at exactly the
Fermi level.  As detailed in the accompanying article by Tahir-Kheli,$^4$
a number of key experimental observations that are otherwise anomalous,
such as the temperature dependence of the NMR spin relaxation rates and
Knight shifts, the Hall effect, the resistivity, and Josephson tunneling,
are easily explained with the resulting band structure and the IBP model.

	What follows is a detailed account of how this Hubbard model 
was developed for $La_{1.85}Sr_{0.15}CuO_4$.  The resulting band structure
is discussed in length.  In addition we find that other copper oxide
superconducting materials can be understood based largely on geometric
constraints.  As an example, we offer simple arguments as to why
$YBa_2Cu_3O_{6+\delta}$ appears to show two characteristic $T_c$'s over
an extended doping range.
\bigskip
\noindent{\bf II. Calculational Details.}

\noindent{\bf II.A.  Cluster Calculations}

Parameters for the Hubbard model were extracted from restricted open-shell
density functional calculations on the cluster $CuO_6$.  These calculations
used the gradient corrected BLYP functional$^5$ with the standard 6-31+G*
basis set$^6$ on the oxygen atoms and Hay and Wadt's$^7$ effective
core potential and basis set on the copper.  All calculations were
performed using the Jaguar$^8$ ab initio electronic structure program on a
dual processor 200 MHz Pentium Pro running Linux.

	The $CuO_6$ cluster was embedded in a point charge array of
1364 ions.  The ions had the formal charges of +2.000 for $Cu$, -1.925
for $O$, -2.000 for apical $O$, and +2.925 for $La/Sr$.  The total
cluster and point charge array had the $D_{4h}$ symmetry of the
tetragonal unit cell and was 5 unit cells wide (18.940 \AA) in the
$a$ and $b$ directions and 3 unit cells tall (39.618 \AA) in the $c$
direction.  Fractional charges were used at the edges.  It should be noted
that at the low temperatures where superconductivity appears, the crystal
shows a $C_{2h}$ distortion to an orthorhombic unit cell.  This should
be a relatively small perturbation, so for the sake of simplicity the
higher symmetry structure was used.  However, we note here that
the distortion will have some important implications in regard to
superconductivity.  This will be discussed below.  The tetragonal crystal
structure was taken from Hazen$^9$ and is given in Table I.

	To obtain the Hubbard parameters, the density of a single
state was optimized and the resulting orbitals were localized.  By then
making specific combinations of the localized orbitals and not allowing
them to relax in subsequent DFT calculations, it was then possible to
determine the Hubbard parameters associated with the localized orbitals.
For example, evaluating the energy of the state where there is one hole
in the localized $Cu\ d_{x^2-y^2}$ orbital ($^2B_{1g}\ [ CuO_6]^{-10}$)
and using the state where there are no holes at all 
($^1A_{1g}\ [CuO_6]^{-11}$) as a reference yields the orbital energy
for $Cu\ d_{x^2-y^2}$.  Evaluating the energy of the state where
there are two holes in the $Cu\ d_{x^2-y^2}$ orbital 
($^1A_{1g}\ [CuO_6]^{-9}$) then leads to the self-Coulomb term for
this orbital.  Similarly, evaluating the energy of a state where there
is one hole in the plus combination of the localized $Cu\ d_{x^2-y^2}$
and $O\ p_{\sigma}$ orbitals as compared to the state where there is
one hole in the minus combination of these orbitals leads to the matrix
element coupling the two orbitals.  Clearly, all principle nearest
neighbor Hubbard parameters describing the set of $Cu$, $O$, and apical
$O$ orbitals can be obtained in such a fashion using just this single
$CuO_6$ cluster.

	The density that was optimized was for the undoped ground state, 
$^2B_{1g}\ [CuO_6]^{-10}$.  This state has one hole in an orbital that
is about 50\% $Cu\ d_{x^2-y^2}$ and 50\% $O\ p_{\sigma}$.  The state
was chosen because it represents the closest approximation to the true
density that can be obtained with this finite cluster.  Note however that
the point charge array reflects a doped state ($x=0.15$) so the total
charge on the system (cluster $+$ point charges) is -0.3.  While it was
not possible to treat the cluster with a fractional charge to make the
total charge on the system neutral, this discrepancy was effectively
removed by the procedure described below in section II.B.1.

	The orbitals were localized using the Pipek-Mezey$^{10}$ localization
procedure which maximizes the sum of the squares of the atomic Mulliken
populations over basis functions.  This procedure was done in several
steps.  First, orbitals within a given irreducible representation were
localized and identified.  This produced localized $Cu\ d_{x^2-y^2}$
($B_{1g}$) and $d_{z^2}$ ($A_{1g}$) orbitals, and symmetry combinations
of the $O\ p_{\sigma}$ ($A_{1g}$, $E_u$, and $B_{1g}$), $O\ p_{\pi}$
($B_{2g}$, $E_u$, and $A_{2g}$), and apical $O\ p_z$ ($A_{1g}$ and
$A_{2u}$) orbitals.  The procedure was then used on the symmetry
combinations of the oxygen orbitals to obtain completely localized 
$O\ p_{\sigma}$, $O\ p_{\pi}$, and apical $O\ p_z$ orbitals.

	Hubbard parameters were derived by evaluating the DFT
energies with fixed orbitals ({\it i.e.} non-SCF) as follows:

\item{1)}  The energy of the $^1A_{1g}\ [CuO_6]^{-11}$ state (having no 
holes) was evaluated.  This was used as our reference state.

\item{2)}  The energies of the doublet $[CuO_6]^{-10}$ states (each
having a single hole in one of the localized orbitals) were evaluated.
This yielded the orbital energies, $E_i^0$.

\item{3)}  The energies of the singlet $[CuO_6]^{-9}$ states (each having 
two holes in one of the localized orbitals) were evaluated.  This yielded
the self-Coulomb repulsion energies, $J_{ii}$.

\item{4)}  The Hartree-Fock energies of the triplet and open-shell
singlet $[CuO_6]^{-9}$ states (each having two holes in different
localized orbitals) were evaluated.  This yielded the exact exchange
energy between orbitals, $K_{ij}$.

\item{5)}  The energies of the triplet $[CuO_6]^{-9}$ states (each having 
two holes in different localized orbitals) were evaluated.  This yielded
the Coulomb repulsion energy between orbitals, $J_{ij}$.

\item{6)}  The energies of the doublet $[CuO_6]^{-10}$ states (each having 
one hole in either the plus or minus combination of pairs of localized
orbitals) were evaluated.  This yielded the matrix elements coupling
pairs of orbitals, $T_{ij}$.  Note that the fully symmetric combinations
of the oxygen orbitals were used in the evaluation of these terms
rather than the completely localized orbitals.

\noindent{Parameters obtained from this procedure are listed in Table II.
This parameter set will be referred to as the unscaled set.}

	The Hubbard model that was developed included explicitly
the $Cu\ d_{x^2-y^2}$ and $d_{z^2}$ orbitals, the $O\ p_{\sigma}$ orbitals
(two per unit cell), and the apical $O\ p_z$ orbitals (two per unit cell).
This led to a total of six bands.  Solving the Hubbard model was done
in two parts.  The first took the input parameters (orbital energies
and coupling terms) and found the single electron energies $\epsilon_n(k)$ 
and wavefunctions $\phi_n(k)$ where n is the band index. These $k$ states 
were filled to the required doping level and the orbital occupations
were evaluated.  In the second step, these orbital occupations were used
to reevaluate the orbital energies.  The procedure was repeated until
self-consistency was achieved.  The Hubbard model is explained in more
detail in the accompanying article.$^4$

The Hubbard model was solved assuming no dispersion in
the $z$ direction.  In calculating the density of states, however,
$z-axis$ dispersion was included perturbatively by assuming a dominant
coupling through the apical $O\ p_z$ orbitals.  Further details follow.
\bigskip
\noindent{\bf II.B.  Orbital Energy Evaluation}

\noindent{\bf II.B.1.  Coulomb Potential}

When orbital occupations are evaluated in the diagonalization step, 
the change in the Coulomb field had to be incorporated into the orbital
energies, $E_i$.  This is done by dividing up this field into that due
to the $CuO_6$ cluster and that due to the point charge array.  Since the
reference orbital energies, $E_i^0$, were originally defined for the
case in which all $CuO_6$ orbitals were doubly occupied ($^1A_{1g}\
[CuO_6]^{-11}$), the change in the orbital energies ($E_i$) due to the
change in the occupation of the $CuO_6$ orbitals ($N_i$) were determined
from,

$$E_i = E_i^0 - (2 - N_i)J_{ii} - \sum_{j{\neq}i} (2-N_j)(J_{ij} - 
{1 \over 2} K_{ij}),\ \ \ \ \ \  N_i > 1 \eqno (1)$$

$$E_i = E_i^0 - J_{ii} - \sum_{j{\neq}i} (2-N_j)(J_{ij} - {1 \over 2} K_{ij}),
\ \ \ \ \ \ \ \ \ \ \ \ \ \ \ \ N_i {\leq} 1 \eqno (2)$$

\noindent where the $Cu\ d_{x^2-y^2}$ and $d_{z^2}$ orbitals, the 
four $O\ p_{\sigma}$ orbitals and the two apical $O\ p_z$ orbitals of
the $CuO_6$ cluster were included in the summation.

	The Coulomb potential due to the changing long range field 
was evaluated by first subtracting off the potential due to the 1364
ion point charge array used in the DFT calculations.  This was evaluated
as a classical point charge Coulomb interaction at each of the $Cu$,
$O$, and apical $O$ sites of the $CuO_6$ cluster.  A similar Coulomb
interaction was evaluated with a larger array having 22374 ions (17 unit
cells wide in the $a$ and $b$ directions (64.396 \AA) and 5 unit cells
tall in the $c$ direction (66.030 \AA)).  This was done to improve the
long range Coulomb field over that which was used in the cluster
calculations.  This new Coulomb field was also broken up into components
due to the $Cu$, $O$, apical $O$, and $La/Sr$ sites.  These fields were
then appropriately scaled based on the orbital occupations from the
Hubbard model and the effect was incorporated into the new orbital
energies.
\bigskip
\noindent{\bf II.B.2.  Correlation}

An important point should be made here about the effect of correlation.  
As can be seen in equations (1-2), the self Coulomb interaction ($J_{ii}$)
is treated separately from the other Coulomb interactions.  This is a
deviation from the mean field approximation of Hartree-Fock theory and
conventional LDA band structure calculations.  The mean field
approximation would instead use the orbital energy correction equation:

$$E_i = E_i^0 - \sum_j (2-N_j)(J_{ij} - {1 \over 2} K_{ij}) \eqno(3)$$
				 
\noindent While equation (3) may be adequate for many materials, it
breaks down in the limit of weakly interacting particles where the self
Coulomb terms are much larger than the coupling matrix elements ($J_{ii}
>> T_{ij}$).  As can be seen from the data in Table II, this is the
case for $La_{1.85}Sr_{0.15}CuO_4$.  In this regime, equations (1-2)
become valid.  The difference between equations (1-2) and (3) can be
seen when one orbital is at half occupancy ($N_i=1$).  In equation (2)
the orbital energy is lowered by the full $J_{ii}$ term while in equation
(3) the orbital energy is only lowered by ($J_{ii} - {1 \over 2}
K_{ii}$) or equivalently ${1 \over 2} J_{ii}$.  Equations (1-2)
assumes that correlation localizes all spins, while equation (3) 
inappropriately assumes the system has ionic character.  With equation
(3) there is a tendency to completely empty a band before electrons
are removed from a second band in order to minimize the self-Coulomb
energy.  With the proper treatment of $J_{ii}$ in equations (1-2) there
is instead a tendency to remove electrons from multiple bands in order
to minimize the Coulomb repulsions between different orbitals. 

	Variations on the type of correlation expressed in equations 
(1-2) have been introduced by several authors$^{11,12}$ in studies of
$La_2CuO_4$.  Application of the self-interaction correction, or
equivalently spin-polarization, has been successful in describing the
antiferromagnetic state of this material.  As will be seen in this work,
the introduction of correlation through equations (1-2) is critical to
obtaining a band crossing in $La_{1.85}Sr_{0.15}CuO_4$.
\bigskip
\noindent{\bf II.B.3.  Orbital Relaxation}

Perhaps the main limitation in the method used here to obtain the 
Hubbard parameters is that the orbitals were not allowed to relax for
different states.  This will result in Coulomb interactions which are
too high and orbital energies which are too low.  To determine the
effect that orbital relaxation would have on our Hubbard parameters,
we looked at isolated $Cu$, $O$, and apical $O$ atoms in the full point
charge array.  We derived orbital energies and Coulomb interactions for
these lone atoms using both fixed orbitals and fully optimized orbitals.
The Coulomb energies were found to uniformly scale as 0.7 when relaxation
was introduced.  The orbital energies (defined here as $E(\phi^2) -
E(\phi) - J_{jj}$) did not scale quite as uniformly, so different scales
were used for $Cu$, $O$, and apical $O$.  We found that the orbital
energy for $Cu$ scaled as 0.6, for $O$ as 0.8, and for apical $O$ as 0.7.
We applied all these scales to the Hubbard parameters listed in Table
II to produce the set listed in Table III.  It is this corrected set
that was used in our final calculations.  We should point out that we
experimented with some other scaled sets (such as that obtained from 
uniformly scaling both the orbital energies and Coulomb terms by 0.7)
and found no qualitative changes in the band structure.  Even when the
unscaled set was used, the same basic features were still observed.
\bigskip
\noindent {\bf III. Results.}

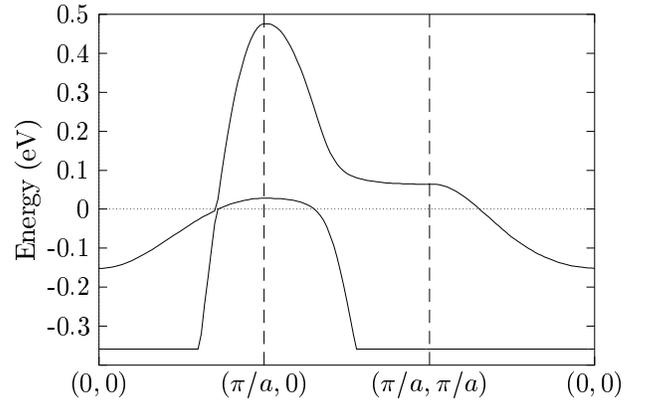
In Figure 2 we show the 2-D dispersion of the top two bands as obtained
with our Hubbard model using the scaled parameter set of Table III.  The
band structure was computed for optimally doped $La_{2-x}Sr_xCuO_4$,
where $x=0.15$.  This doping level corresponds to the removal of a total
of 1.15 electrons per unit cell from the $Cu$/$O$/apical $O$ bands
(undoped $La_2CuO_4$ has 1 electron per unit cell removed from these
bands).  The orbital energies and occupations that were computed
self-consistently from the model are given in Table IV.

	As can be seen in Figure 2, two bands appear to be important 
near the Fermi level.  While no holes have been created in the lower
band the Fermi level is just 0.035 eV above the top of this band.
More importantly, a rigorous crossing, which is critical to the
proposed theory, is seen just 0.153 eV below the Fermi level along 
the $(0,0)-(\pi/a,\pi/a)$ symmetry line (note, there are actually four
crossing points in the full 2-D Brillouin zone at $(k,k)$, $(k,-k)$,
$(-k,k)$, and $(-k,-k)$).  The close proximity of this crossing to the
Fermi level, as determined from this simple model, is a significant
finding.  Given errors in the Hubbard parameters and missing electronic
effects, it is not difficult to conclude at this point that a band
crossing, required of the IBP model, indeed appears to occur at the 
Fermi level of the optimally doped superconductor.  This crossing can
be characterized as arising between a broad ``$B_{1g}$'' band dominated
by $Cu\ d_{x^2-y^2}$ character and a narrower ``$A_{1g}$'' band dominated
by $Cu\ d_{z^2}$ character.

	As shown in Figure 3, it is clearly the case that this band 
crossing is only observed when the self-Coulomb term is treated properly.
Using the mean-field equation (3) we find only a single isolated band
having ``$B_{1g}$'' $Cu\ d_{x^2-y^2}$ character at the Fermi level.  Other
bands are well buried.  We clarify this by noting that ligand field
theory properly predicts that the $Cu\ d_{x^2-y^2}$ orbital is the
most unstable and $d_{z^2}$ is the next most unstable in the Jahn-Teller
distorted octahedron of $CuO_6$.  The highest energy band orbitals of our
Hubbard model reflect that this is true.  However, as electrons are
removed from the $d_{x^2-y^2}$ orbital the energy of this orbital stabilizes
with respect to the $d_{z^2}$ orbital as a result of reduction in the
self-Coulomb energy.  This makes it more favorable to remove electrons
from the $d_{z^2}$ orbital.  Conversely, as electrons are removed from
the $d_{z^2}$ orbital the energy of this orbital stabilizes with respect
to the $d_{x^2-y^2}$ orbital.  Yet because the self-Coulomb energy of the
$d_{z^2}$ orbital is smaller than that of the $d_{x^2-y^2}$ orbital due
to $s-d_{z^2}$ hybridization, it is actually possible to remove {\it more}
electrons from the $d_{z^2}$ orbital than the $d_{x^2-y^2}$ orbital.  Such
considerations are important since they lead directly to the observed
band crossing.

	To take our calculations one step further, we considered what 
electronic effects might be missing that would have the greatest effect
on the position of the crossing point relative to the Fermi level.
Besides errors that might be present in the basic Hubbard parameters
due to basis set limitations, cluster size, and the DFT method itself,
a number of missing key electronic effects can be identified.  All
are expected to lead to only minor perturbations of the band structure,
but their cumulative effect could have an impact on the position of
the band crossing.  These effects include:

\item{1)}  Explicit inclusion of z-axis dispersion instead of the 
perturbative approach taken here.

\item{2)}  Inclusion of additional Hubbard parameters.  Additional 
parameters should all be $<$ 0.04 eV.

\item{3)}  Explicit inclusion of other bands.  Small mixings with other 
bands, in particular the $Cu\ d_{xy}/O\ p_{\pi}$ band, could have an
effect on the position of the crossing.

\item{4)}  Use of more realistic charges in the cluster calculations.

\item{5)}  Inclusion of additional spin correlation.  The current model 
only includes the self-interaction correction to account for the tendency
of weakly interacting systems to localize spin.  However, spin couplings
between different orbitals, in particular the triplet coupling between
$Cu\ d_{x^2-y^2}$ and $d_{z^2}$, has been ignored.

\item{6)}  Explicit inclusion of the $La$ and $Sr$ ions.  The current 
model treats these atoms as point charges having the formal charges +3.0
and +2.0 , respectively.  In reality, these ions are likely less highly
charged.  They also have spatial extent which leads to Pauli repulsions.

\item{7)}  Inclusion of the orthorhombic distortion.  The crystal in 
the superconducting phase is distorted from its high symmetry tetragonal
($D_{4h}$) structure to a lower symmetry orthorhombic ($C_{2h}$)
structure.$^9$  This is manifested by a tilting of the $CuO_6$
units or equivalently a buckling of the $CuO_2$ planes.  An important
effect from this distortion is that two of the four crossing points
(at $(k,-k)$ and $(-k,k)$) become strictly avoided due to the reduced
symmetry. However, it should be stressed that the other two crossing
points (at $(k,k)$ and $(-k,-k)$) are maintained, which is critical to
the proposed IBP theory.

\noindent While it is not clear that all of these effects will be
favorable in terms of moving the band crossing toward the Fermi level,
their combined effect could easily lead to such a change.  We stress
that a perturbation of only 0.153 eV (a small quantity on the chemical
scale) is necessary to observe a Fermi level crossing.

	We have, in fact, applied a specific perturbation to our model in 
order to incorporate one of the above effects.  Investigations of the
effect of $La$ and $Sr$ using a variety of clusters at the BLYP level
showed significant mixing of the $O\ p_{\pi}$ orbitals with the orbitals
of these metals.  The implication of these results is that there is
diffusion of the $O\ p_{\pi}$ electrons onto the $La/Sr$ sites which
has been ignored by the $CuO_6$ cluster calculations.  The effect of
this charge diffusion is to lower the energy of both the $O\ p_{\sigma}$
and $O\ p_{\pi}$ orbitals and raise the energy of the apical $O\ p_z$
orbitals relative to the $Cu$ orbitals. To account for this effect we
included in our Hubbard model calculations an adjustable parameter which
defined the extent of charge transfer from $O\ p_{\pi}$ to $La/Sr$.
While we do not explicitly include these $\pi$ bands in the final model,
Coulomb terms for the $O\ p_{\pi}$ orbitals were evaluated and included
in Tables II and III ($La$ and $Sr$ were still treated as classical
point charges).  This allowed us to include the effect of charge transfer
by altering the Coulomb field in a fashion similar to that explained
in section II.B.1.  Orbital energies were reevaluated to reflect the
change in the Coulomb field due to this charge transfer.

	The primary effect of this charge transfer is to stabilize 
the $\pi$ bands relative to the Fermi level by significantly lowering
the orbital energy of $O\ p_{\pi}$.  A secondary effect, however, is to
raise the crossing point of the two bands of interest closer to the
Fermi level for the optimally doped system.  This was accomplished when
the charge transfer term was empirically adjusted to the value of 0.50
electrons transfered.  That is, the charge on the $La/Sr$ sites was
+2.425 compared to the formal charge of +2.925.  This should be considered
within the range of reasonable charges for these ions (note the charges
on the other atoms are +1.563 for $Cu$, -1.429 for $O$, and -1.777 for
apical $O$).  We should note though, that the extent of charge transfer
calculated here may be an overestimate in light of the fact that the
other electronic effects listed above have not yet been incorporated.
But, as detailed in the accompanying article by Tahir-Kheli,$^4$ the
resulting band structure proves to have the necessary features to explain
a number of key experiments.

	The 2-D dispersion of the top two bands from these calculations 
is shown in Figure 4 and the optimized orbital energies and occupations 
are given in Table V.  In addition the Fermi surfaces are shown in Figure
5 and the density of states of the top two bands is shown in Figure 6.
\bigskip
\noindent{\bf IV. Discussion.}

\noindent{\bf IV.A Band Structure of $\bf La_{1.85}Sr_{0.15}CuO_4$}

	The band crossing we observe arises from the following considerations.
In this discussion, we will refer to the two bands that cross as the
``$B_{1g}$'' and ``$A_{1g}$'' bands.  The crossing produces two new
bands that touch, referred to as $U$(pper) and $L$(ower).  The dispersion
of the ``$B_{1g}$'' band, dominated by $Cu\ d_{x^2-y^2}$ character,
is rather broad, on the order of 2 eV, producing a low density of states.
The dispersion of the ``$A_{1g}$'' band, dominated by $Cu\ d_{z^2}$
character, is in contrast rather narrow, on the order of 0.3 eV, producing 
a high density of states.  At the ($\pi/a,\pi/a$) point, the higher
energy band is ``$B_{1g}$'' in nature, corresponding to the completely
antibonding combination of $Cu\ d_{x^2-y^2}$ and $O\ p_{\sigma}$ orbitals.
The lower energy band is ``$A_{1g}$'' in nature, corresponding to the
antibonding combination of the $Cu\ d_{z^2}$, $O\ p_{\sigma}$, and apical
$O\ p_z$ orbitals.  At the $(0,0)$ point, however, the higher energy
band is ``$A_{1g}$'' in nature and the lower energy band is ``$B_{1g}$.
Due to this change in the relative energetics between the top of both
bands and the bottom of both bands, the ``$B_{1g}$'' and ``$A_{1g}$''
bands must cross, producing a $U$ band and $L$ band which each have
both ``$B_{1g}$'' and ``$A_{1g}$'' character.  This crossing of bands
is strictly avoided everywhere except for one point in the 2-D Brillouin
zone.  Along the $(0,0)-(\pi/a,\pi/a)$ symmetry line, the crossing
is rigorously allowed since the ``$B_{1g}$'' and ``$A_{1g}$'' orbitals
cannot mix along this direction.  Due to this rigorous crossing, the $U$
band and $L$ band must touch.  While the position of this crossing
point (or touching point) is subject to variation, its existence is
quite robust over a wide range of model parameters.

	In the closeup of the density of states shown in Figure 6, 
it can be seen how the crossing of the ``$B_{1g}$'' and ``$A_{1g}$''
bands affects the nature of the $U$ and $L$ bands.  The $U$ band starts
at +0.47 eV and represents the ($\pi/a,\pi/a$) point of the ``$B_{1g}$''
band.  The density of states of this band remains consistently low until
about +0.10 eV where there is a sharp peak.  This peak represents the
change in character of the band from ``$B_{1g}$'' to ``$A_{1g}$'' in the
vicinity of the ($\pi/a,0$) point.   The density of states of the $U$
band remains relatively high at energies below this point, being dominated
by ``$A_{1g}$'' character.  The band terminates at -0.25 eV, which
represents the $(0,0)$ point of the ``$A_{1g}$ band.  The onset of the
$L$ band is characterized by a sharp peak in the density of states at
+0.03 eV.  This peak occurs in the vicinity of the ($\pi/a,\pi/a$)
point of the ``$A_{1g}$'' band.  The peak comes down to a low density
of states near the Fermi level and the density of states remains low,
being dominated by ``$B_{1g}$'' character until about -0.20 eV when
character from other orbitals (in particular the ``$A_{2u}$''
antibonding combination of the apical $O\ p_z$ orbitals) starts to mix in.
The change in character of this band at lower energies should have no
effect on the issue of the band crossing or the IBP model.

The character of these bands can be seen more clearly in Figure 7.
In Figure 7a, the $U$ band is shown to be dominated by $Cu\ d_{x^2-y^2}$
character above 0.10 eV while the $L$ band is dominated by $d_{x^2-y^2}$
below the Fermi level.  In Figure 7b, the $U$ is shown to be dominated
by $Cu\ d_{z^2}$ character below 0.10 eV while the $L$ band is dominated
by $d_{z^2}$ only at the Fermi level.

	While most band structure calculations have only shown a single
band at the Fermi level,$^{13}$ we argue that this is due to the lack
of correlation.  In fact, the work of Shiraishi, {\it et al.},$^{11}$
which included the effect of spin-polarization, found that doping of
$La_2CuO_4$ resulted in the formation of two types of holes from two
distinct bands.  While no band crossing was noted, the two types of
holes were characterized as being $Cu\ d_{x^2-y^2}$ and $d_{z^2}$.  Eto,
{\it et al.}$^{14}$ also suggested the importance of the $Cu\ d_{z^2}$
orbital based on cluster model calculations.  We find that this work
lends support to the findings reported here.
\bigskip
\noindent{\bf IV.B. Band Structure of Related Materials}

	Certain qualitative features in the density of states for 
$La_{2-x}Sr_xCuO_4$ ($LASCO$) should be characteristic of many
superconducting copper oxide materials, such as $YBa_2Cu_3O_{6+\delta}$
($YBCO$), $Bi_2Sr_2Ca_{n-1}Cu_nO_{2n+6+y}$,
$Tl_2Ba_2Ca_{n-1}Cu_nO_{2n+4+\delta}$, and others.  Each should be
characterized by a peak in the $U$ band just above the top of the $L$
band.  This is due to band repulsions away from the $(0,0)-(\pi/a,\pi/a)$
diagonal which introduce ``$A_{1g}$'' character into the $U$ band.
When these band repulsions are large enough (such as they are
at ($\pi/a,0$)), the peak in the density of states of the $U$ band should
occur above the top of the $L$ band.  The $L$ band, on the other hand,
should have a sharp peak at its onset at ($\pi/a,\pi/a$) which vanishes
as the band becomes ``$B_{1g}$'' in character.

	Although the electronic structure of $YBCO$ should share
common features with $LASCO$, it differs in one important regard:
the local symmetry of $YBCO$ in the superconducting state is $D_{2h}$ as
compared to $C_{2h}$ for $LASCO$.$^9$  This symmetry will not allow a rigorous
crossing of the two bands along the diagonal (or any other point) in
the 2-D Brillouin zone.  However, it should be recognized that the dual
$CuO_2$ planes of $YBCO$ lead to four bands in contrast to the two bands
of $LASCO$.  Dispersion of these four bands in the $z$ direction leads to
two bands having ``$A_g$'' symmetry at $k_z=0$ and $k_z=\pi/c$ and two
bands having ``$B_{1u}$'' symmetry.  The two ``$A_g$'' bands are
principally composed of the bonding combinations of the $Cu\ d_{x^2-y^2}$
orbitals from the two planes and the bonding combinations of the $Cu\ d_{z^2}$
orbitals from the two planes.  The ``$B_{1u}$'' bands are the antibonding
analogues.  While the two ``$A_g$'' bands are not precluded from mixing
at any symmetry point, and the two ``$B_{1u}$'' bands are also not
precluded from mixing, the ``$A_g$'' and ``$B_{1u}$'' bands cannot mix when
$k_z=0$ or $k_z=\pi/c$.  This suggests that crossings could occur
between the ``$A_g$'' bands and the ``$B_{1u}$'' bands.  Based on this
idea, we propose the following scenario.  (Note, we do not address the
effects of the chain $CuO$ bands.  These bands will perturb the planar
$CuO_2$ bands, but we suspect the essential topology described below
holds.)

	There are two sets of two bands ($U$ and $L$) having qualities
similar to those shown here for $LASCO$.  One set has ``$A_g$'' symmetry
while the other has ``$B_{1u}$'' symmetry.  The primary difference between
these bands and those of $LASCO$ is that the crossing along the
$(0,0)-(\pi/a,\pi/a)$ direction is avoided.  Since the $Cu\ d_{x^2-y^2}$
dispersion in the $z$ direction is expected to be small, the $d_{x^2-y^2}$
components of these bands (analogous to the ``$B_{1u}$'' bands of
$LASCO$) should be nearly degenerate.  In contrast, the $Cu\ d_{z^2}$
components of these bands (analogous to the ``$A_{1g}$'' bands of $LASCO$)
should be separated in energy.

	The onset of superconductivity in $YBCO$ should occur when the 
Fermi level coincides with a band crossing.  This likely occurs at $k_z=\pi/c$
between the ``$B_{1u}$'' $L$ band (which is completely antibonding in
the $z$ direction) and the ``$A_g$'' $U$ band.  Interestingly, it is
not required to occur along the $(0,0)-(\pi/a,\pi/a)$ direction.  In fact,
there is likely a double crossing within a quadrant of the Brillouin zone,
as depicted in Figure 8.  In this schematic, we view the Fermi surfaces
as squared circles centered around $(0,0)$ for ``$B_{1u}$'' $L$ and
``$A_g$'' $L$ and centered around $(\pi/a,\pi/a)$ for ``$B_{1u}$'' $U$.
Superconductivity begins when the ``$A_g$'' $L$ and ``$B_{1u}$'' $U$
bands first touch.  Since the radius of the ``$B_{1u}$'' $U$ Fermi
surface is increasing faster than the radius of the ``$A_g$'' $L$ Fermi
surface is decreasing (due to the difference in their densities of
states), two crossing points can be sustained over a wide doping range.
At sufficiently higher doping levels, a second set of band crossings of
a similar nature should occur at $k_z=0$.  This proposal easily explains
the extended doping range observed for $YBCO$ and the appearance of two
$T_c$'s in different doping regimes for this material.  Furthermore,
it suggests a reason for the increasing $T_c$ for $LASCO$ vs. underdoped
$YBCO$ vs. optimally doped $YBCO$.  We correlate an increase in the
number of crossing points, or, more correctly, the number of crossing
points which are thermally accessible, to an increase in $T_c$.  Indeed,
it is observed that superconductivity begins with $YBa_2Cu_3O_{6.6}$
showing a $T_c$ of 60 K.  This $T_c$ is sustained upon further doping
until a rapid increase to $T_c=90$ K is observed near $YBa_2Cu_3O_{6.9}$.
Further doping to $YBa_2Cu_3O_{7.0}$ maintains $T_c$ at this higher
temperature.$^{15}$

	Similar analysis can be applied to the bismuth$^{16}$ and
thallium$^{17}$ systems.  We note that in these systems an increase in
$T_c$ is correlated with an increase in the number of $CuO_2$ planes
per unit cell.  Following the above arguments this makes sense in that
it leads to more bands which produce more band crossings as shown
schematically in Figure 9.  For a three plane system we are now dealing
with six bands.  Four of these bands will be symmetric with respect to
reflection through the middle plane (bands $g1$, $g2$, $g3$, and $g4$)
and two of these bands will be antisymmetric (bands $u1$ and $u2$).
We argue that band repulsions near the ($\pi/a,0$) point, as seen in
the $LASCO$ band structure, would be stronger for the $g$ bands, since
$d_{z^2}$ character should appear in these bands at a higher energy than
in the $u$ bands.  This could produce a crossing between the $u1$ and $g2$
bands at $k_z=0$ and $k_z=\pi/c$ as depicted in the figure.  Since the
two surfaces might be expected to be highly coincident at this point,
the number of thermally accessible crossing points should be high.
Clearly, it can be seen that the addition of more $CuO_2$ planes increases
the probability of favorable crossing situations as illustrated here, and this
should lead to potentially higher $T_c$'s.

	Finally we wish to note that the electron doped system, 
$Nd_{2-x}Ce_{x}CuO_4$,$^{18}$ may be substantially different in regard to the
nature of the two bands that cross.  While we anticipate one of the
bands will be ``$B_{1g}$'' $Cu\ d_{x^2-y^2}$ in character, the other 
band is likely not ``$A_{1g}$'' $Cu\ d_{z^2}$ since this band only
appears as electrons are removed from the system.  We suggest instead
that the second band is a $Nd/Ce$ band.  It follows from this suggestion
that high temperature superconductivity is not dependent on the specific
crossing of the $Cu\ d_{x^2-y^2}$ and $d_{z^2}$ bands.  We stress, in
fact, the only requirement of the IBP model that we now see is that some
sort of crossing of bands occurs at the Fermi level.  This leads us to
be optimistic that with the careful exploitation of symmetry, entirely
new classes of high temperature superconductors will be developed in
our future.
\bigskip
\noindent {\bf V. Conclusion.}

We have presented the results of a new Hubbard model calculation on 
the optimally doped superconducting material $La_{1.85}Sr_{0.15}CuO_4$.
We conclude from these calculations that there is a crossing of two
bands which occurs at the Fermi level.  One of these bands is ``$B_{1g}$'' in
character, dominated by $Cu\ d_{x^2-y^2}$ and the other is ``$A_{1g}$'' 
in character, dominated by $Cu\ d_{z^2}$.  The crossing rigorously occurs
along the $(0,0)-(\pi/a,\pi/a)$ symmetry line of the 2-D Brillouin zone.
As detailed in the accompanying article by Tahir-Kheli,$^4$ an inter-band
pairing (IBP) of electrons between these two bands leads to
superconductivity.  It can only occur at the critical doping level where
the Fermi energy coincides with the band crossing, {\it i.e.} $x=0.15$.
The resulting density of states from this work is used in the accompanying
work to explain a number of key experiments on this material.

	Extension of the band model obtained for $LASCO$ in these 
calculations to $YBCO$ provides an easy explanation for the observation
of a wide doping range for superconductivity to occur in this material
as well as an explanation for the observation of two $T_c$'s in different
doping regimes.  We anticipate at this point that other materials will be
similarly understood largely through geometric considerations.
\bigskip
\noindent{\bf Acknowledgment:}  The authors wish to thank Dr. Jean-Marc 
Langlois for many useful discussions.
\vfill
\eject
\noindent{\bf References.}

\item{$^1$}J.G. Bednorz and K.A. M\"uller, Z. Phys. B {\bf 64}, 189 (1986).

\item{$^2$}J. Tahir-Kheli, in {\it Proceedings of the 10th Anniversary
HTS Workshop on Physics, Materials and Applications}, ed. B. Batlogg,
C.W. Chu, W.K. Chu, D.U. Gubser, and K.A. M\"uller (World Scientific,
New Jersey: 1996), 491-492.

\item{$^3$}H. Takagi, R.J. Cava, M. Marezio, B. Batlogg, J.J. Krajewski,
W.F.Peck, Jr., P. Bordet, D.E. Cox, Phys. Rev. Lett. {\bf 68}, 3777 (1992).

\item{$^4$}J. Tahir-Kheli, http://www.firstprinciples.com, cond-mat/9711170

\item{$^5$}J.C. Slater, {\it Quantum Theory of Molecules and Solids,
Vol. 4: The Self-Consistent Field for Molecules and Solids} (McGraw-Hill,
New York, 1974); A. D. Becke, Phys. Rev. A {\bf 38}, 3098 (1988); C. Lee,
W. Yang, and R. G. Parr, Phys. Rev. B {\bf 37}, 785 (1988); implemented as
described in B. Miehlich, A. Savin, H. Stoll, and H. Preuss,
Chem. Phys. Lett. {\bf 157}, 200 (1989).

\item{$^6$}W.J. Hehre and J.A. Pople, J. Chem. Phys. {\bf 56}, 4233 (1972).

\item{$^7$}P. J. Hay and W. R. Wadt, J. Chem. Phys. {\bf 82}, 299 (1985).  

\item{$^8$}M. N. Ringnalda, J.-M. Langlois, R. B. Murphy, B. H. Greeley,
C. Cortis, T. V. Russo, B.  Marten, R. E. Donnelly, Jr., W. T. Pollard,
Y. Cao, R. P. Muller, D. T. Mainz, J. R.  Wright, G. H. Miller,
W. A. Goddard III, and R. A. Friesner, Jaguar (formerly PS-GVB) v2.3,
Schr\"ošdinger, Inc., 1996.

\item{$^9$} R.M. Hazen, in {\it Physical Properties of High Temperature
Superconductors II}, ed. D.M. Ginsberg (World Scientific, New Jersey; 1990),
121-198.

\item{$^{10}$}J. Pipek and P. G. Mezey, J. Chem. Phys. {\bf 90}, 4916 (1989).

\item{$^{11}$}K. Shiraishi, A. Oshiyama, N. Shima, T. Nakayama, and H.
Kamimura, Solid State Comm. {\bf 66}, 629 (1988).

\item{$^{12}$}A. Svane, Phys. Rev. Lett. {\bf 68}, 1900 (1992).

\item{$^{13}$}W.E. Pickett, Rev. Mod. Phys. {\bf 61}, 433 (1989), and
references therein.

\item{$^{14}$}M. Eto, R. Saito, and H. Kamimura, Solid State Comm.
{\bf 71}, 425 (1989).

\item{$^{15}$}R.J. Cava, B. Batlogg, C.H. Chen, E.A. Rietman, S.M.
Zahurak, and D. Werder, Nature {\bf 329}, 423 (1987).

\item{$^{16}$}C. Michel, M. Hervieu, M.M. Borel, A. Grandin, F. Deslandes,
J. Provost, and B. Raveau, Z. Phys. B {\bf 68} 421 (1987); H. Maeda,
Y. Tanaka, M. Fukutomi, and T. Asano, Jpn. J.  Appl. Phys. {\bf 27},
L209 (1988).

\item{$^{17}$}Z.Z. Sheng and A.M. Hermann, Nature {\bf 332}, 55 (1988); 
Z.Z. Sheng and A.M. Hermann {\bf 332}, 138 (1988).

\item{$^{18}$}Y. Tokura, H. Takagi, and S. Uchida, Nature {\bf 337}, 345
(1989).
\vskip2.0truein
\noindent{\bf Table I.} Crystal structure of $La_{1.85}Sr_{0.15}CuO_4$ 
(in \AA).

\vskip 0.3truein
\halign{\noindent#\hfil &\qquad \hfil#\hfil \cr
\noalign{\bigskip\hrule\smallskip}
\noalign{\hrule\medskip}
$a$ & 3.788 \cr
$b$ & 3.788 \cr
$c$ & 13.206 \cr
\noalign{\medskip\hrule\medskip}
$Cu$ & (0.000,0.000,0.000) \cr
$O(1)$ & (0.500,0.000,0.000) \cr
$O(2)$ & (0.000,0.000,0.182) \cr
$La/Sr$ & (0.000,0.000,0.361) \cr
\noalign{\medskip\hrule\smallskip}
\noalign{\hrule\bigskip}}
\bigskip
\noindent{\bf Table II.} Unscaled Hubbard parameters (in eV).  $E$
is an orbital energy, $T$ an orbital coupling matrix element, $J$
a Coulomb repulsion term, and $K$ an exchange energy term.

\vskip 0.5truein
\halign{\noindent#\hfill &\quad \hfill# &\qquad #\hfill &\quad \hfill#
&\qquad #\hfill &\quad \hfill# \cr
\noalign{\bigskip\hrule\smallskip}
\noalign{\hrule\medskip}
$E(x^2-y^2)$ & -3.109 & $J(x^2-y^2/z^2)$ & 25.563 & 
$K(x^2-y^2/z^2)$ & 1.219 \cr
$E(z^2)$ & -3.338 & $J(x^2-y^2/O\ p_{\sigma})$ & 7.999 & 
$K(x^2-y^2/O\ p_{\sigma})$ & 0.081 \cr
$E(O\ p_{\sigma})$ & -10.417 & $J(x^2-y^2/O2\ p_z)$ & 6.471 & 
$K(x^2-y^2/O2\ p_z)$ & 0.028 \cr
$E(O2\ p_z)$ & -12.442 & $J(x^2-y^2/O\ p_{\pi})$ & 7.053 &
$K(x^2-y^2/O\ p_{\pi})$ & 0.006 \cr
$J(x^2-y^2/x^2-y^2)$ & 29.281 & $J(z^2/O\ p_{\sigma})$ & 6.471 & 
$K(z^2/O\ p_{\sigma})$ & 0.028 \cr
$J(z^2/z^2)$ & 25.990 & $J(z^2/O2\ p_z)$ & 6.855 &
$K(z^2/O2\ p_z)$ & 0.262 \cr
$J(O\ p_{\sigma}/O\ p_{\sigma})$ & 17.375 & $J(z^2/O\ p_{\pi})$ & 6.813 &
$K(z^2/O\ p_{\pi})$ & 0.008 \cr
$J(O2\ p_z/O2\ p_z)$ & 11.365 & $J(O\ p_{\sigma}/O\ p_{\sigma}')$ & 5.345 &
$K(O\ p_{\sigma}/O\ p_{\sigma}')$ & 0.035 \cr
$T(x^2-y^2/O\ p_{\sigma})$ & 1.347 & $J(O\ p_{\sigma}/O\ p_{\sigma}'')$ & 
3.890 & $K(O\ p_{\sigma}/O\ p_{\sigma}'')$ & 0.008 \cr
$T(z^2/O\ p_{\sigma})$ & 0.514 & $J(O\ p_{\sigma}/O2\ p_z)$ & 5.144 &
$K(O\ p_{\sigma}/O2\ p_z)$ & 0.211 \cr
$T(z^2/O2\ p_z)$ & 1.076 & $J(O\ p_{\sigma}/O\ p_{\pi})$ & 14.299 &
$K(O\ p_{\sigma}/O\ p_{\pi})$ & 0.677 \cr
$T(O\ p_{\sigma}/O\ p_{\sigma}')$ & 0.368 & $J(O\ p_{\sigma}/O\ p_{\pi}')$ &
5.548 & $K(O\ p_{\sigma}/O\ p_{\pi}')$ & 0.092 \cr
$T(O\ p_{\sigma}/O\ p_{\sigma}'')$ & -0.041 & $J(O\ p_{\sigma}/O\ p_{\pi}'')$ &
3.953 & $K(O\ p_{\sigma}/O\ p_{\pi}'')$ & 0.008 \cr
$T(O\ p_{\sigma}/O2\ p_z)$ & 0.078 & $J(O2\ p_z/O2\ p_z')$ & 3.502 &
$K(O2\ p_z/O2\ p_z')$ & 0.098 \cr
$T(O2\ p_z/O2\ p_z')$ & 0.493 & $J(O2\ p_z/O\ p_{\pi})$ & 4.868 &
$K(O2\ p_z/O\ p_{\pi})$ & 0.087 \cr
\noalign{\medskip\hrule\smallskip}
\noalign{\hrule\bigskip}}
\vfill
\eject
\noindent{\bf Table III.} Scaled Hubbard parameters (in eV).  $E$
is an orbital energy, $T$ an orbital coupling matrix element, $J$
a Coulomb repulsion term, and $K$ an exchange energy term.

\vskip 0.5truein
\halign{\noindent#\hfill &\quad \hfill# &\qquad #\hfill &\quad \hfill#
&\qquad #\hfill &\quad \hfill# \cr
\noalign{\bigskip\hrule\smallskip}
\noalign{\hrule\medskip}
$E(x^2-y^2)$ & 1.063 & $J(x^2-y^2/z^2)$ & 17.894 & 
$K(x^2-y^2/z^2)$ & 1.219 \cr
$E(z^2)$ & 0.596 & $J(x^2-y^2/O\ p_{\sigma})$ & 5.599 & 
$K(x^2-y^2/O\ p_{\sigma})$ & 0.081 \cr
$E(O\ p_{\sigma})$ & -10.071 & $J(x^2-y^2/O2\ p_z)$ & 4.532 & 
$K(x^2-y^2/O2\ p_z)$ & 0.028 \cr
$E(O2\ p_z)$ & -8.709 & $J(x^2-y^2/O\ p_{\pi})$ & 4.937 &
$K(x^2-y^2/O\ p_{\pi})$ & 0.006 \cr
$J(x^2-y^2/x^2-y^2)$ & 20.497 & $J(z^2/O\ p_{\sigma})$ & 5.318 & 
$K(z^2/O\ p_{\sigma})$ & 0.028 \cr
$J(z^2/z^2)$ & 18.193 & $J(z^2/O2\ p_z)$ & 4.799 &
$K(z^2/O2\ p_z)$ & 0.262 \cr
$J(O\ p_{\sigma}/O\ p_{\sigma})$ & 12.163 & $J(z^2/O\ p_{\pi})$ & 4.769 &
$K(z^2/O\ p_{\pi})$ & 0.008 \cr
$J(O2\ p_z/O2\ p_z)$ & 7.956 & $J(O\ p_{\sigma}/O\ p_{\sigma}')$ & 3.742 &
$K(O\ p_{\sigma}/O\ p_{\sigma}')$ & 0.035 \cr
$T(x^2-y^2/O\ p_{\sigma})$ & 1.347 & $J(O\ p_{\sigma}/O\ p_{\sigma}'')$ & 
3.601 & $K(O\ p_{\sigma}/O\ p_{\sigma}'')$ & 0.008 \cr
$T(z^2/O\ p_{\sigma})$ & 0.514 & $J(O\ p_{\sigma}/O2\ p_z)$ & 3.601 &
$K(O\ p_{\sigma}/O2\ p_z)$ & 0.211 \cr
$T(z^2/O2\ p_z)$ & 1.076 & $J(O\ p_{\sigma}/O\ p_{\pi})$ & 10.009 &
$K(O\ p_{\sigma}/O\ p_{\pi})$ & 0.677 \cr
$T(O\ p_{\sigma}/O\ p_{\sigma}')$ & 0.368 & $J(O\ p_{\sigma}/O\ p_{\pi}')$ &
3.884 & $K(O\ p_{\sigma}/O\ p_{\pi}')$ & 0.092 \cr
$T(O\ p_{\sigma}/O\ p_{\sigma}'')$ & -0.041 & $J(O\ p_{\sigma}/O\ p_{\pi}'')$ &
2.767 & $K(O\ p_{\sigma}/O\ p_{\pi}'')$ & 0.008 \cr
$T(O\ p_{\sigma}/O2\ p_z)$ & 0.078 & $J(O2\ p_z/O2\ p_z')$ & 2.451 &
$K(O2\ p_z/O2\ p_z')$ & 0.098 \cr
$T(O2\ p_z/O2\ p_z')$ & 0.493 & $J(O2\ p_z/O\ p_{\pi})$ & 3.408 &
$K(O2\ p_z/O\ p_{\pi})$ & 0.087 \cr
\noalign{\medskip\hrule\smallskip}
\noalign{\hrule\bigskip}}
\bigskip
\noindent{\bf Table IV.} Computed orbital energies and orbital occupations
for optimally doped $La_{1.85}Sr_{0.15}CuO_4$ taken from our Hubbard model
using the scaled parameter set of Table III and no $O\ p_{\pi}$ to $La/Sr$
charge transfer.  All energies are relative to the Fermi level (in eV).

\vskip 0.3truein
\halign{\noindent#\hfill &\quad \hfill#\hfill &\quad \hfill#\hfill \cr
\noalign{\bigskip\hrule\smallskip}
\noalign{\hrule\medskip}
Orbital & Energy & Occupation \cr
\noalign{\medskip\hrule\medskip}
$Cu\ x^2-y^2$ & -2.570 & 1.572 \cr
$Cu\ z^2$ & -1.663 & 1.785 \cr
$O\ p_{\sigma}$ & -4.254 & 1.806 \cr
$O2\ p_z$ & -1.611 & 1.941 \cr
\noalign{\medskip\hrule\smallskip}
\noalign{\hrule\bigskip}}

\vfil\eject
\noindent{\bf Table V.} Computed orbital energies and orbital occupations
for optimally doped $La_{1.85}Sr_{0.15}CuO_4$ taken from our Hubbard model
using the scaled parameter set of Table III and 0.50 electron $O\ p_{\pi}$
to $La/Sr$ charge transfer.  All energies are relative to the Fermi level
(in eV).

\vskip 0.3truein
\halign{\noindent#\hfill &\quad \hfill#\hfill &\quad \hfill#\hfill \cr
\noalign{\bigskip\hrule\smallskip}
\noalign{\hrule\medskip}
Orbital & Energy & Occupation \cr
\noalign{\medskip\hrule\medskip}
$Cu\ x^2-y^2$ & -2.403 & 1.770 \cr
$Cu\ z^2$ & -2.092 & 1.666 \cr
$O\ p_{\sigma}$ & -6.122 & 1.929 \cr
$O2\ p_z$ & -0.852 & 1.777 \cr
\noalign{\medskip\hrule\smallskip}
\noalign{\hrule\bigskip}}
\vfill\eject
\end